\documentstyle[multicol,prl,aps,epsfig]{revtex}
\begin{document}
\draft
\title{Scattering of Bunched Fractionally Charged Quasiparticles}
\author{Y.~C.~Chung, M.~Heiblum, and V.~Umansky}
\address{Braun Center for Submicron Research, Department of Condensed
 Matter Physics, Weizmann Institute of Science, Rehovot 76100, Israel}
\date{\today}
\maketitle

\begin{abstract}
The charge of fractionally charged quasiparticles, proposed by
Laughlin to explain the fractional quantum Hall effect (FQHE), was
recently verified by measurements.  Charge $q=e/3$ and $e/5$ ($e$
is the electron charge), at filling factors $\nu=1/3$ and $2/5$,
respectively, were measured.  Here we report the unexpected
\textit{bunching} of fractional charges, induced by an extremely
weak backscattering potential at exceptionally low electron
temperatures ($T<10~mK$) - deduced from shot noise measurements.
Backscattered charges $q=\nu e$, specifically, $q=e/3$, $q=2e/5$,
and $q<3e/7$, in the respective filling factors, were measured.
For the same settings but at an only slightly higher electron
temperature, the measured backscattered charges were $q=e/3$,
$q=e/5$, and $q=e/7$. In other words, \textit{bunching} of
backscattered quasiparticles is taking place at sufficiently low
temperatures. Moreover, the backscattered current exhibited
distinct temperature dependence that was correlated to the
backscattered charge and the filling factor.  This observation
suggests the existence of 'low' and 'high' temperature
backscattering states, each with its characteristic charge and
characteristic energy.
\end{abstract}

 \noindent

\begin{multicols}{2}
While Laughlin's argument, explaining the fractional quantum Hall
effect (FQHE)~\cite{r1}, is useful in predicting the charge of the
quasiparticles for fractional filling factors of the type
$\nu=1/(2p+1)$, the composite Fermion (CF) model~\cite{r2} is
helpful in more general filling factors, such as
$\nu=n/(2np+1)$,with $p$ and $n$ integers.  The predicted
quasiparticle charge is always $e^*=e/(2np+1)$.  For $p=1$, $2$,
and $3$ and $n=1$, or alternatively for $\nu=1/3$, $2/5$ and
$3/7$, we expect $q=e/3$, $e/5$ and $e/7$, respectively.  Indeed
recent quantum shot noise measurements confirmed these predictions
at $\nu=1/3$ and $2/5$.  The shot noise, in turn, resulting from
weak backscattering of quasiparticles by a quantum point contact
(QPC), was measured at electron temperatures $30 \sim 80~mK$, led
to charges $e/3$ at $\nu=1/3$~\cite{r3} and $e/5$ at
$\nu=2/5$~\cite{r4} - as expected. Here we report on shot noise
measurements in the extreme limits of: (a) weak backscattering ($r
\sim 2\%$), where backscattering events are so rare assuring their
independence, and (b) extremely low electron temperatures
($T_{min} \sim 9~mK$).  In this regime of a barely perturbed
electron system we measured, surprisingly, shot noise
corresponding to backscattered charges $q=\nu e$, namely, $q=e/3$,
$2e/5$ and $\sim 3e/7$ at $\nu=1/3$, $2/5$ and $3/7$,
respectively. In other words, backscattering in this regime is
that of correlated $p$ quasiparticles.

Measurements were conducted in a high mobility low-density
two-dimensional electron gas (2DEG), embedded in a GaAs-AlGaAs
heterojunction.  The magnetic field was set well within the
conductance plateaus of the FQHE.  For example, the magnetic field
at $\nu=1/3$ was $B \sim 14.26~T$ - near the center of the
$g_Q=e^2/3h$ plateau (Fig. 1(a)).  A QPC type potential, induced
in the 2DEG with two biased metallic gates deposited on the
surface of the heterojunction, served as a controlled
backscattering potential. A multiple-terminal-configuration (Fig.
1(b)) was employed in order to keep the input and output
differential conductance constant, $g=g_Q$ - independent of the
transmission of the QPC~\cite{r5}. The differential conductance
was measured with a 3~Hz AC, 0.5~$\mu$V RMS, excitation voltage
superimposed on a DC bias that was restricted to the linear regime
of the QPC. The spectral density of the noise, $S$, was measured
as function of DC current at a center frequency 1.4~MHz and
bandwidth $\sim$30~kHz (determined by a LC resonant circuit; see
Refs. 3 and 4 for more details).  A low noise cryogenic
preamplifier, in the vicinity of the sample, amplified the voltage
fluctuations in terminal A, followed by an amplifier and a
spectrum analyzer at ambient temperature, measuring the RMS
fluctuations at 1.4~MHz. The temperature of the electrons was
determined by measuring the equilibrium noise, $S=4k_B Tg$, with
$k_B$ the Boltzman constant. Shot noise was determined by
subtracting the current independent noise from the total noise
signal.

Figure 1(c) shows typical differential conductance curves of a QPC
at bulk filling $\nu=1/3$.  Measurements were conducted at the
lowest electron temperatures $T \sim 9$~mK for different
backscattering potential strengths (controlled by the QPC gates
voltage $V_g$). Even a relatively weak backscattering potential,
with high voltage transmission $t=g/g_Q \sim 0.7$ ($r\sim 0.3$),
lead to rather strong backscattering near zero applied voltage.
Moreover, both the voltage and temperature dependence of the
differential conductance were positive - qualitatively agreeing
with the prediction of the chiral Luttinger liquid (CLL)
model~\cite{r6,r7,r8,r9}. However, when the QPC potential was
tuned even weaker this dependence reversed sign (see Figs. 1(c)
and 1(d)). We concentrate now on the limiting case, namely, the
extremely weak backscattering regime, with the temperature
dependence of the backscattered current for $r \sim 0.03$ shown in
Fig. 1(d).  A distinct positive slope over a decade of the current
is seen in the $log(I_B)$ vs. $log(T)$ characteristic, with $I_B$
the backscattered current.

\begin{figure}
\begin{center}
\leavevmode \epsfxsize=8.5 cm  \epsfbox{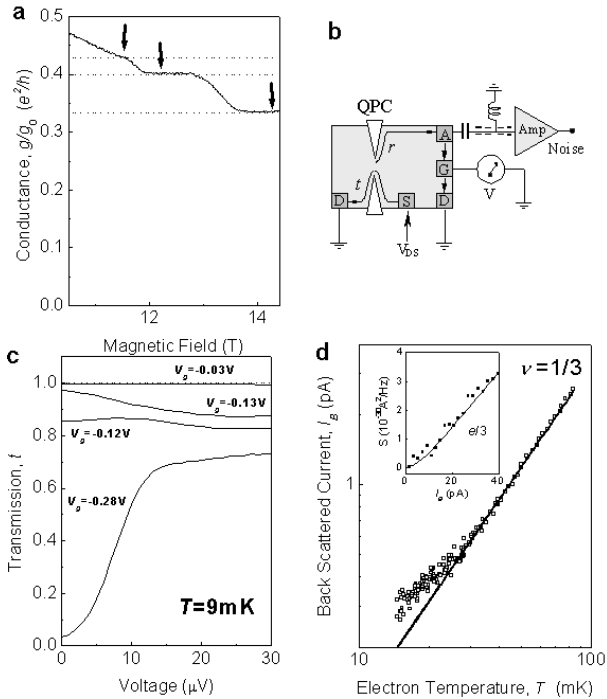}
\end{center}
\vspace{0 cm} \caption{\textbf{(a)} Quantum Hall conductance as a
function of magnetic field.  Filling factors were established by
the pointed out magnetic fields.  Similar results were obtained at
different magnetic fields around the middle of the conductance
plateau. \textbf{(b)} The measurement set up of shot noise and
differential conductance. The noise generated by the QPC passed
through a resonant circuit tuned to 1.4~HMz and amplified by a
cryogenic amplifier.  This small capacitance at A allowed only the
high frequency component through.  The multiple-terminal geometry
kept the conductance seen from S and A constant.  \textbf{(c)}
Typical dependence of the transmission coefficient on bias voltage
for different QPC gate voltage, at $\nu=1/3$.  When the QPC is
very weakly pinched off ($V_g=-0.03$~V) the transmission has a
very weak negative dependence on the applied bias - oppositely to
a CLL.  \textbf{(d)} The backscattered current as a function of
electron temperature with AC 10~$\mu$V RMS is applied. The curve
can be fitted with a single slope.  Inset: The shot noise
generated by a very weakly pinched off QPC ($t \sim 0.97$) at a
filling factor $\nu=1/3$ and electron temperature of 9~mK.  Noise
is classical and quasiparticle charge is $e/3$.}
\end{figure}

We turn now to shot noise measurements.  The low temperature
quantum shot noise of partitioned particles in the CLL regime was
predicted~\cite{r6,r7} and later found~\cite{r8,r9} to be highly
non-classical (non-Poissonian).  This is expected since
backscattering of quasiparticles is correlated and energy
dependent.  However, when backscattering events are very rare and
the temperature is finite, it is expected that scattering events
are stochastic with a resultant classical-like shot
noise~\cite{r10}. Indeed the measured spectral density of the shot
noise, $S$, shown in the inset of Fig. 1(d), is classical-like.
The solid line is the expected shot noise due to stochastic
scattering of independent particles at $T=9$~mK and charge
$q=e/3$~\cite{r11}. It depends on $V$, $q$, $t$, and $T$ via
$S=4k_BTg+2qI_Bt \cdot \Theta (T,V)$, with
$\Theta(T,V)=coth(qV/(2k_BT))-2k_BT/(qV)$, $I_B=Vg_Q(1-t)$, and
$g_Q=\nu e^2/h$. Here, for $qV \gg k_BT$, $\Theta(T,V) \sim 1$ and
the dependence of $S$ on $I_B$ is linear, and for $qV \ll k_BT$
the Johnson-Nyquist thermal noise dominates. Note that at $T \sim
0$ and $t \rightarrow 1$, as in our case, $S \simeq 2qI_B$. The
excellent agreement between experiment and prediction proves that
scattering events of $e/3$ quasiparticles are independent down to
the lowest temperatures provided that the backscattering potential
is extremely weak.

\begin{figure}
\begin{center}
\leavevmode \epsfxsize=8.5 cm  \epsfbox{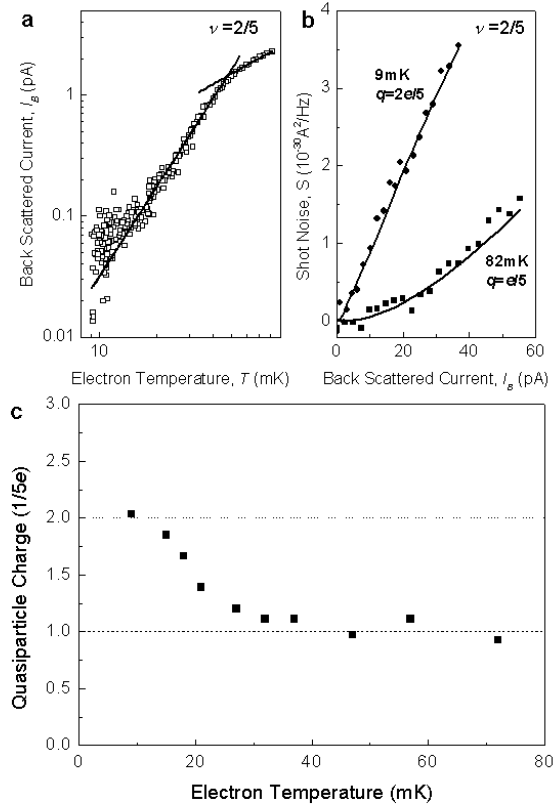}
\end{center}
\vspace{0 cm} \caption{\textbf{(a)} Backscattered current as
function of the electron temperature at a filling factor
$\nu=2/5$.  Two distinct slopes are observed with a transition
temperature of about 45mK. \textbf{(b)} Shot noise at two
different temperatures. The backscattered quasiparticle charge is
$2e/5$ at 9~mK and $e/5$ at 82~mK. The QPC was set to reflect some
2\% of the impinging current at the two temperatures. \textbf{(c)}
The temperature dependence of the scattered charge when QPC was
set to reflect some 2\% of the impinging current.}
\end{figure}

We now study the regime of extremely weak backscattering at $p=2$,
namely, electron filling factor $\nu=2/5$.  The magnetic field was
tuned to $B=12.2$~T within the $g_Q=(2/5)e^2/h$ plateau (see Fig.
1(a)) and the QPC to $r \sim 0.02$.  Note that the general
features seen in Fig. 1(c) at $\nu=1/3$ are also found at
$\nu=2/5$. We measured the temperature dependence of the
backscattered current, as shown in Fig. 2(a), and find this time
two distinct slopes in $log(I_B)$ vs. $log(T)$ - with a crossover
at $T \sim 45$~mK.  We then measured the shot noise at different
electron temperatures and found it, again, to be classical-like in
all temperatures (see in Fig. 2(b) two extreme examples).  When
determining the charge in a most general filling factor one has to
rely on the CF model. According to that model the reflected
current, carrying the noise, is that of CFs in the $2^{nd}$ Landau
level (LL), namely $p=2$, with the $e/3$ quasiparticles (in the
$1^{st}$ LL) being fully transmitted without contributing to the
shot noise~\cite{r4}.  Hence, one can define an effective
transmission coefficient of $2^{nd}$ LL CFs, $t_{eff}=(t \cdot
g_{2/5}-g_{1/3})/(g_{2/5}-g_{1/3})=6t-5$, which is smaller than
the bare transmission $t$. However, when $t$ is very close to
unity $t_{eff} \sim t$, and the determination of $q$ is not
sensitive to the exact value of $t$.  The two solid lines in Fig.
2(b), agreeing with the data, are the calculated shot noise
according to the expression above with charges $q=2e/5$ at $T \sim
9$~mK and $q=e/5$ at $T \sim 82$~mK (with electron temperatures
determined independently).  While the scattered charge at
\textit{high} temperature $q=e/5$, had been verified
before~\cite{r4}, the scattered charge at \textit{low} temperature
$q=\nu e=2e/5$, was unexpected. Figure 2(d) shows the charge
evolution as the temperature is being increased in the range
$9$~mK$<T<50$~mK. Most of the change takes place over a 20~mK
range.  In comparison, temperature dependence measurements were
conducted in a separately patterned Hall bar. While the $\nu=1/3$
conductance plateau remained unaffected at this temperature range
and the longitudinal resistance $R_{xx}$ increased with
temperature, but with dependence quite different than that of
$I_B$ from the QPC.

Does such unexpected \textit{bunching} take place also in higher
CF filling factors $p=3$, namely, at $\nu=3/7$?  Because of the
relatively weak magnetic field at the $\nu=3/7$ (B=11.5~T) the
many-body energy gap required to establish the $g_Q=(3/7)e^2/h$
plateau is rather small. Consequently, the plateau is barely
established even at the lowest temperature (see Fig. 2(a)) and
there is a finite backscattered current through the bulk (some
0.5\%) and the minimum longitudinal resistance $R_{xx}>0$.
Inducing a very weak QPC potential in the 2DEG increased the
backscattered current and produced a measurable shot noise, as
seen in Fig. 3. Using the effective transmission (that is more
sensitive here to the bare $t$) the fitted charge at the lowest
temperature was extremely sensitive to minute variations in the
electron temperature and seems to hover in the range $(2 \sim
2.5)e/7$. Warming the electrons to $T \geq 27$~mK lowered
significantly the shot noise and established firmly a
quasiparticle charge of $e/7$.  This is the first measurement of
such small fractional charge.  The higher scattered charge at the
lowest temperature indicates again bunching of $e/7$
quasiparticles - very much like the behavior at $\nu=2/5$,
however, it seems that an even lower temperature than our lowest
temperature ($T<9$~mK) is needed to establish bunching of three
$e/7$ quasiparticles to a charge $q=3e/7$ as well as to achieve a
perfect FQH plateau.

\begin{figure}
\begin{center}
\leavevmode \epsfxsize=8.5 cm  \epsfbox{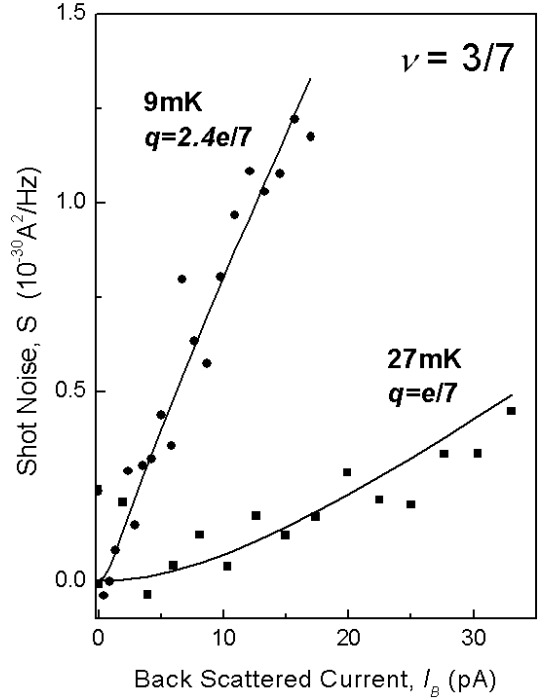}
\end{center}
\vspace{-0.5 cm} \caption{Shot noise at a filling factor $\nu=3/7$
at two different temperatures. The backscattered quasiparticle
charge is found to be around $(2 \sim 2.5)e/7$ at 9~mK and $e/7$
at 27~mK.  The QPC was set to reflect some 2\% of the impinging
current at the two temperatures.}
\end{figure}

Summarizing our results one should recall: (a) the 2DEG is rather
pure with mobility $2\times 10^6 ~cm^2 V^{-1} s^{-1}$, hence,
scattering is dominated by the weak potential of the QPC; (b) the
electron temperature is very low ($T \sim 9$~mK, $k_BT \sim
0.8$~$\mu$eV), minimizing thermal noise and alleviating any
ambiguity in analyzing the data; (c) the QPC is very open, leading
to very rare backscattering events; (d) shot noise is
classical-like with linear dependence of noise on current -
suggesting independent scattering of quasiparticles with a
specific charge; (e) pinching the QPC ever so slightly more
renders both the DC current and shot noise to be highly non-linear
functions of voltage - suggesting correlated scattering of
quasiparticles with charge dependent on bias.  These results
confirm, that in a barely perturbed 2DEG and $T \sim 0$, a very
weakly bound scattering state is formed, with transport dominated
by independent scattering events of $p$ bunched quasiparticles
with charge $q=\nu e$.  In the CF model, rare backscattering
events of simultaneous $p$ quasiparticles, one from each LL, are
taking place.

Relying on the CLL model, Kane and Fisher predicted such
possibility of bunching due to backscattering via to a point
scatterer~\cite{r12}. However, their expression for the
backscattered current,$I_B \propto \upsilon^2 T^{-|\alpha|}$  ,
with $\upsilon$ an energy independent backscattering amplitude and
$\alpha$ a coefficient depended on the scattered charge, suggests
a decrease of the backscattered current with temperature -
contradicting our data. Note though that since our QPC is almost
fully open, its potential is expected to be rather smooth and
shallow with energy dependent backscattering amplitude,
$\upsilon=\upsilon(T, V)$. This dependence might dominate the
behavior of the backscattered current leading to our result.
Still, we stress, that our observations were reproducible among
samples with different QPCs and different cooling cycles; hence,
we believe that it is not sensitive to the details of the QPC
potential.

It should be noted that our observed bunching at low temperatures
is quite different from the already observed bunching by strong
backscattering potentials (at $\nu=1/3$, bunching of
quasiparticles leads to \textit{electron}
scattering)~\cite{r5,r8,r13}. In the latter case the FQHE state
does not exist in the barrier region, hence preventing the
existence of elementary quasiparticles - \textit{forcing} the
quasiparticles to bunch to an electron.  Here, however, the FQHE
state is hardly perturbed in the barrier region, still allowing
the existence of elementary quasiparticles.  Hence, the
\textit{spontaneous} bunching of $p$ quasiparticles is possibly
related to their fractional statistics, namely, their partly
Bosonic nature, encouraging them to bunch upon scattering.

We thank C. Kane and A. Stern for helpful discussions and D.
Mahalu for the submicron lithography. The work was partly
supported by the Israeli Academy of Science and by the
German-Israeli Foundation (GIF).

\end{multicols}

\end{document}